\documentclass{article}
\usepackage{amsmath}
\usepackage{graphicx}

\def\be{\begin{equation}}
\def\ee{\end{equation}}
\def\bea{\begin{eqnarray}}
\def\eea{\end{eqnarray}}

\def\<{\langle}
\def\>{\rangle}

\def\CG{{\cal G}}

\def\CM{{\cal M}}

\def\GeV{\text{\ GeV}}
\def\ce{\varepsilon}

\def\la{\lambda}
\def\La{\Lambda}

\def\cross#1#2{[\vec #1\!\times\!\vec #2]}
\let\oldbar\bar
\def\bar{\; \oldbar}

\begin{document}
\title{Two-photon exchange in nonrelativistic approximation}
\author{Dmitry~Borisyuk$^1$, Alexander~Kobushkin$^{1,2}$\\[2mm]
\it\small $^1$Bogolyubov Institute for Theoretical Physics, 14-B Metrologicheskaya street, Kiev 03680, Ukraine\\
\it\small $^2$National Technical University of Ukraine "Igor Sikorsky KPI", 37 Prospect Peremogy, Kiev 03056, Ukraine}
\maketitle
\abstract{
We calculate two-photon exchange amplitudes for the elastic electron-hadron scattering in the non\-relativistic approximation, and obtain analytical formulae for them. Numerical calculations are performed for proton and $^3$He targets. Comparing our numerical results with relativistic calculations, we find that the real part of the amplitude is described well at moderate $Q^2$, but the imaginary part strongly differs from the relativistic result. Thus the nonrelativistic approximation should not be used for calculation of observables which depend on the imaginary part of the amplitude, such as single-spin asymmetries.
}

\section{Introduction}

It is well-known that, at very small electron energies, two-photon exchange (TPE) amplitudes approach the limit which corresponds to the scattering off a point particle \cite{McKinley}.
On the other hand, at high energies, fully relativistic calculations of TPE on proton exist, taking into account either elastic \cite{BMT, ourDisp} or elastic and inelastic \cite{ourPiN} intermediate states.
Somewhere in between a nonrelativistic approximation (NRA) should work: the proton still can be considered nonrelativistic, but already cannot be considered point-like (the electron, because of its small mass, is always ultrarelativistic), but this question is poorly studied in the literature.
Nevertheless, for example, when studying TPE on nuclei \cite{KobDeu, KobTri}, full relativistic treatment is usually impossible,
and one has to resort, in some sense, to NRA --- e.g. using nonrelativistic nuclear wavefunction.
The question then arises, what is the error of such an approximation and the limits of its applicability.

In the present work we wish to study, what will result if we apply NRA for the elastic TPE contribution on the proton. Since for this quantity full relativistic calculation exists, we can compare the results and thus learn the area of NRA validity. The equations obtained here can also be useful in other situations where proton is nonrelativistic (for example is a part of a nucleus described by nonrelativistic wavefunction). We also apply our formulae to the $^3$He nucleus (considered as a single particle), which has the same spin and parity as the proton.

In Ref.~\cite{Lewis} potential scattering in the second Born approximation was studied, which correspond to real part of TPE correction the $F_1$ form factor. In the present work we will calculate, in NRA, all three invariant TPE amplitudes (real an imaginary parts) and compare with the results of the previous works.

The paper is organaized as follows. In Sec.~II the analytical equations for the TPE ampltiudes in NRA are derived.
Numerical results are given in Sec.~III and conclusions in Sec.~IV. There are two appendices with technical details.

\section{Equations for the TPE amplitudes}

\subsection{One photon exchange}
We use usual notation for kinematics, where $k$ ($k'$) and $p$ ($p'$) are initial (final) electron and proton momenta, 
$u$ ($u'$) and $U$ ($U'$) are electron and proton spinors, and momentum transfer is $q = p'-p = k-k'$.

At first, let us consider the one-photon exchange:
\be
  \CM_1 = -4\pi\alpha \; j_\mu \bar U' \Gamma_\mu U
\ee
where $\alpha$ is fine structure constant, $j_\mu = \bar u' \gamma_\mu u$ is leptonic current, $j_\mu q_\mu = 0$, and
\be\label{Gamma}
  \Gamma_\mu(q) = F_1(q^2) \gamma_\mu - \frac{1}{4M} [\gamma_\mu, \hat q] F_2(q^2)
\ee
Here $\hat q \equiv \gamma_\mu q_\mu$, $M$ is proton mass and $F_1$ and $F_2$ are Dirac and Pauli form factors of the proton.
For the convenience we include the denominator of the photon propagator into the form factors:
\be
  F_i(q^2) \to \frac{F_i(q^2)}{q^2 - \lambda^2 + i0}
\ee
where $\lambda$ is fictitious infinitesimal photon mass.

Writing down proton spinors in NRA as
\be
 U(p) = \sqrt{2M} \left( {w \atop \frac{\vec p\vec \sigma}{2M} w} \right)
\ee
where $w$ is two-component spinor, $w^+ w = 1$,
and leaving out terms of 2nd and higher order in $p$, $p'$, we obtain
\be\label{M1gamma}
  \CM_1 = -4\pi\alpha \left\{ F_1 (2M j_0 - \vec p_+ \vec j) S_0 + i F_m \cross{j}{q} \vec S \right\}
\ee
where $F_m = F_1 + F_2$ is magnetic form factor, $S_0 = w'^+w$, $\vec S = w'^+\vec \sigma w$ (thus $\vec S$ is twice the matrix element of the proton spin).
Further we will need an expression analogous to (\ref{M1gamma}) for the general-case $pp\gamma$ vertex:
\be\label{Gamma2}
  \Gamma_\mu = F_1 \gamma_\mu - \frac{1}{4M} [\gamma_\mu, \hat q] F_2 + \frac{p_{+\mu}}{4M^2} \hat k_+ F_3
\ee
where $p_+ = p+p'$, $k_+ = k + k'$.
Additional term, proportional to $F_3$, can be easily obtained by substitution $j_\mu \to j_0 \frac{k_{+\mu}}{2M}$ and $F_1,F_m \to F_3$ in Eq.~(\ref{M1gamma}), and the amplitude becomes
\be\label{Mgeneric}
  \CM = -4\pi\alpha \left\{ 2M j_0 S_0 \left( F_1 + \CG_3 \right) + i \cross{j}{q} \vec S \, F_m + \frac{i}{2k} j_0 \cross{k_+}{q} \vec S \, \CG_3 \right\}
\ee
where $\CG_3 = \frac{p_+k_+}{4M^2} F_3 \approx \frac{|\vec k|}{M} F_3$, as in Ref.~\cite{ourDisp}
(note that the term with $\vec p_+ \vec j$ from Eq.(\ref{M1gamma}) goes beyond our accuracy and thus was left out).
\subsection{Two photon exchange}
Now let us consider the TPE amplitude:
\be\label{M2gamma}
% \CM = -i\frac{(4\pi\alpha)^2}{(2\pi)^4} \int d^4q \bar U' \Gamma_\nu(q_2) \frac{\hat p''+M}{p''^2-M^2+i0} \Gamma_\mu(q_1) U
% \left\{ \bar u' \gamma_\nu \frac{\hat k'' + m}{k''^2 - m^2 + i0} \gamma_\mu u + 
%         \bar u' \gamma_\mu \frac{\hat{\tilde k}'' + m}{\tilde k''^2 - m^2 + i0} \gamma_\nu u \right\} = \nonumber \\
% = -i\frac{(4\pi\alpha)^2}{(2\pi)^4} \int d^4q \bar U' \Gamma_\nu(q_2) \frac{\hat p''+M}{p''^2-M^2+i0} \Gamma_\mu(q_1) U \left\{ L_{\nu\mu}(k-q_1) + L_{\mu\nu}(k-q_2) \right\}
 \CM_2 = -i\frac{(4\pi\alpha)^2}{(2\pi)^4} \int d^4q_1 \bar U' \Gamma_\nu(q_2) \frac{\hat p+\hat q_1+M}{(p+q_1)^2-M^2+i0} \Gamma_\mu(q_1) U \left\{ L_{\nu\mu}(k-q_1) + L_{\mu\nu}(k-q_2) \right\}
\ee
where $q_2 = q - q_1$,
\be
  L_{\nu\mu}(k'') = \bar u' \gamma_\nu \frac{\hat k'' + m}{k''^2 - m^2 + i0} \gamma_\mu u
\ee
and $m$ is electron mass.
The $\Gamma_\mu$ vertex for the off-shell proton is written in the form (\ref{Gamma}), because this ensures gauge invariance.

We will consider a kinematical region where all momenta are much greater that electron mass, but much less the nucleon mass:
\be
  m \ll p,k,q_1,q_2 \ll M
\ee
Note that thus we will have
\be\label{q10+q20}
 q_{10} + q_{20} = q_0 = \frac{\vec p'^2}{2M} - \frac{\vec p^2}{2M} \approx 0
\ee
At first, expand the denominator of the proton propagator:
\be
  %\frac{1}{p''^2 - M^2 + i0} = 
  \frac{1}{(p+q_1)^2 - M^2 + i0} \approx \frac{1}{(M + q_{10})^2 - (\vec p + \vec q_1)^2 - M^2 + i0} \approx \frac{1}{2M} \; \frac{1}{q_{10} + i0}
\ee
Now expand the numerator, denoting the leptonic part for brevity as %$j_{2\nu} j_{1\mu}$.
\be\label{leptonic}
 \tilde L_{\nu\mu} = L_{\nu\mu}(k-q_1) + L_{\mu\nu}(k-q_2)
\ee
We obtain:
%\bea%\label{M2gamma-jj}
%  \frac{1}{2M} \bar U' \Gamma_\nu(q_2) (\hat p+\hat q_1+M) \Gamma_\mu(q_1) U j_{2\nu} j_{1\mu}
%  = F_1(q_1^2) F_1(q_2^2) \, 2M j_{10} j_{20} S_0 +\nonumber \\
%  + i F_1(q_1^2) F_1(q_2^2) \, q_{10} \cross{j_2}{j_1} \vec S
%  + i F_1(q_1^2) F_m(q_2^2) \, j_{10} \cross{j_2}{q_2} \vec S
%  + i F_1(q_2^2) F_m(q_1^2) \, j_{20} \cross{j_1}{q_1} \vec S
%\eea
\bea\label{M2gamma-jj}
  \bar U' \Gamma_\nu(q_2) \frac{\hat p+\hat q_1+M}{(p+q_1)^2-M^2+i0} \Gamma_\mu(q_1) U \tilde L_{\nu\mu}
  = \frac{1}{2M} \; \frac{1}{q_{10} + i0} \left\{ 2M F_1(q_1^2) F_1(q_2^2) \, S_0 \tilde L_{00} +\nonumber \right. \\
  \left.
  + i F_1(q_1^2) F_1(q_2^2) \, S_i q_{10} \ce_{ijk} \tilde L_{jk}
  + i F_1(q_1^2) F_m(q_2^2) \, S_i \ce_{ijk} q_{2k} \tilde L_{j0}
  + i F_1(q_2^2) F_m(q_1^2) \, S_i \ce_{ijk} q_{1k} \tilde L_{0j} \right\}
\eea
(where $i$, $j$, $k$ denote spatial components). Taking into account (\ref{q10+q20},\ref{leptonic}) we find that the expression in curly brackets is symmetrical with respect to exchange $q_1 \leftrightarrow q_2$, therefore we may change
\be
  \frac{1}{q_{10}+i0} \to \frac{1}{2}\left( \frac{1}{q_{10}+i0} + \frac{1}{q_{20}+i0} \right)
  = \frac{1}{2}\left( \frac{1}{q_{10}+i0} + \frac{1}{-q_{10}+i0} \right) = -i\pi \delta(q_{10})
\ee
Thus the integration over $q_{10}$ gets trivial, and the second term in (\ref{M2gamma-jj}) vanishes.
The electron propagator takes the form
\be
  %\frac{1}{k''^2 - m^2 + i0} = 
  \frac{1}{(k-q_1)^2 - m^2 + i0} = \frac{1}{2\vec k \vec q_1 - \vec q_1^2 + i0} = - \frac{1}{(\vec k - \vec q_1)^2 - k^2 - i0}
\ee
and similarly
\be
  %\frac{1}{\tilde k''^2 - m^2 + i0} = 
  \frac{1}{(k-q_2)^2 - m^2 + i0} = - \frac{1}{(\vec k - \vec q_2)^2 - k^2 - i0}
\ee
In the last two formulae $k^2$ is understood in the 3-dimensional sense, i.e. $k^2 \equiv \vec k^2$, and $k = |\vec k|$.
The same notation is applied further in the text to all other vectors (thus, e.g. $q^2 = \vec q^2 > 0$).

So,
\bea
  \CM_2 = \frac{2\alpha^2}{\pi} \int  \frac{d^3 q_1}{(\vec k - \vec q_1)^2 - k^2 - i0}
  \left\{ 2M F_1(q_1^2) F_1(q_2^2) \, S_0 \bar u' \gamma_0 (\hat k - \hat q_1) \gamma_0 u + \right. \nonumber \\
  \left.  + i F_1(q_2^2) F_m(q_1^2) \, S_i \ce_{ijk} q_{1k} \bar u' \gamma_0 (\hat k - \hat q_1) \gamma_j u
          + i F_1(q_1^2) F_m(q_2^2) \, S_i \ce_{ijk} q_{2k} \bar u' \gamma_j (\hat k' + \hat q_2) \gamma_0 u
    \right\}
\eea
or, changing in the last term $q_1 \leftrightarrow q_2$
\bea
  \CM_2 = \frac{2\alpha^2}{\pi} \int d^3 q_1
  \left\{ 2M F_1(q_1^2) F_1(q_2^2) \; S_0 \frac{\bar u' \gamma_0 (\hat k - \hat q_1) \gamma_0 u}{(\vec k - \vec q_1)^2 - k^2 - i0} + \right. \nonumber \\
  \left.    + i F_1(q_2^2) F_m(q_1^2) \; S_i \ce_{ijk} q_{1k} 
               \left[ \frac{\bar u' \gamma_0 (\hat k - \hat q_1) \gamma_j u}{(\vec k - \vec q_1)^2 - k^2 - i0}
                     +\frac{\bar u' \gamma_j (\hat k' + \hat q_1) \gamma_0 u}{(\vec k' + \vec q_1)^2 - k^2 - i0} 
               \right]
    \right\}\label{Int-k-origin}
\eea

Applying easy, though long, algebraic transformations to this formula, and comparing the result with (\ref{Mgeneric}) (see Appendix \ref{app:algebraic}), we obtain the following equations for the TPE amplitudes in NRA

\bea\label{Int-k-begin}
  \delta F_1 + \delta \CG_3 &=& \frac{\alpha k q^2}{2 k_+^2} \int d^3q_1 \frac{ F_1(q_1^2) F_1(q_2^2) }{k''^2 - k^2 - i0} \left\{k_+^2 + 2 \vec k_+ \vec k''\right\} \\
  \delta F_m              &=& 2 \alpha k \int d^3q_1 \frac{ F_m(q_1^2) F_1(q_2^2)}{k''^2 - k^2 - i0}
                                    \left\{ k''^2 + \frac{q^2}{4} - \vec q \vec k'' - \frac{(\vec k_+ \vec k'')^2}{k_+^2} \right\} \\
  \delta \CG_3            &=& \frac{2\alpha k}{k_+^2} \int d^3q_1 \frac{ F_m(q_1^2) F_1(q_2^2)}{k''^2 - k^2 - i0}
                                    \left\{ \frac{4k^2+q^2}{k_+^2} (\vec k_+ \vec k'')^2 + (\vec q \vec k'')^2 - 4k^2 k''^2  \right\}
\label{Int-k-end}
\eea
where $\vec k'' = \vec k - \vec q_1$ and $\vec q_2 = \vec q - \vec q_1$, and the prefix $\delta$ indicates TPE contribution to the corresponding generalized form factor. For performing the integrations see Appendix \ref{app:integrals}.

Eq.~(\ref{Int-k-begin}) is in agreement with the results of Ref.~\cite{Lewis}.

\section{Numerical results}

\begin{figure}[h]
\includegraphics[width=0.33\textwidth]{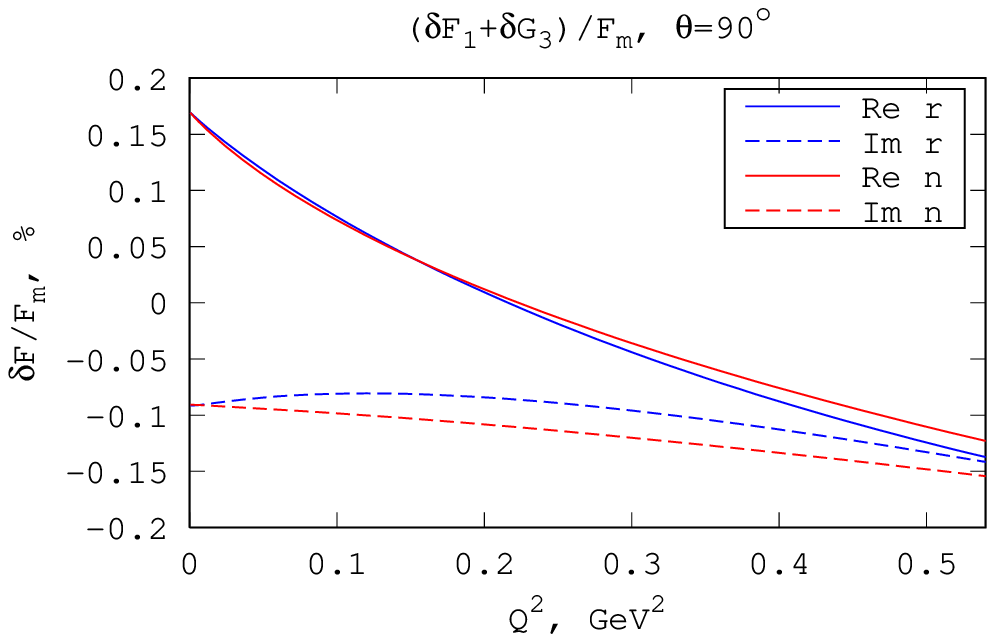}\hfill
\includegraphics[width=0.33\textwidth]{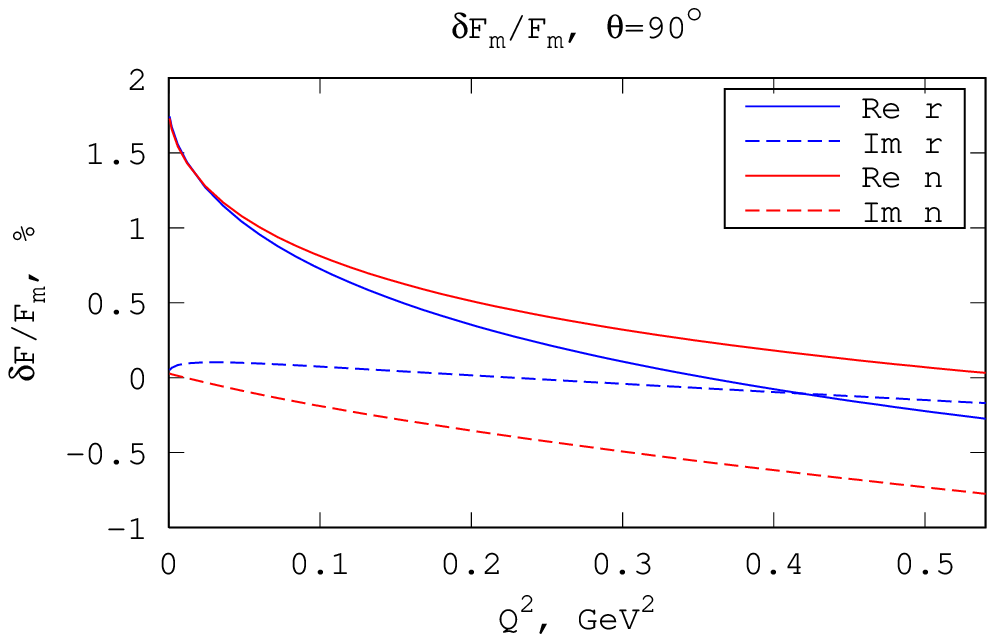}\hfill
\includegraphics[width=0.33\textwidth]{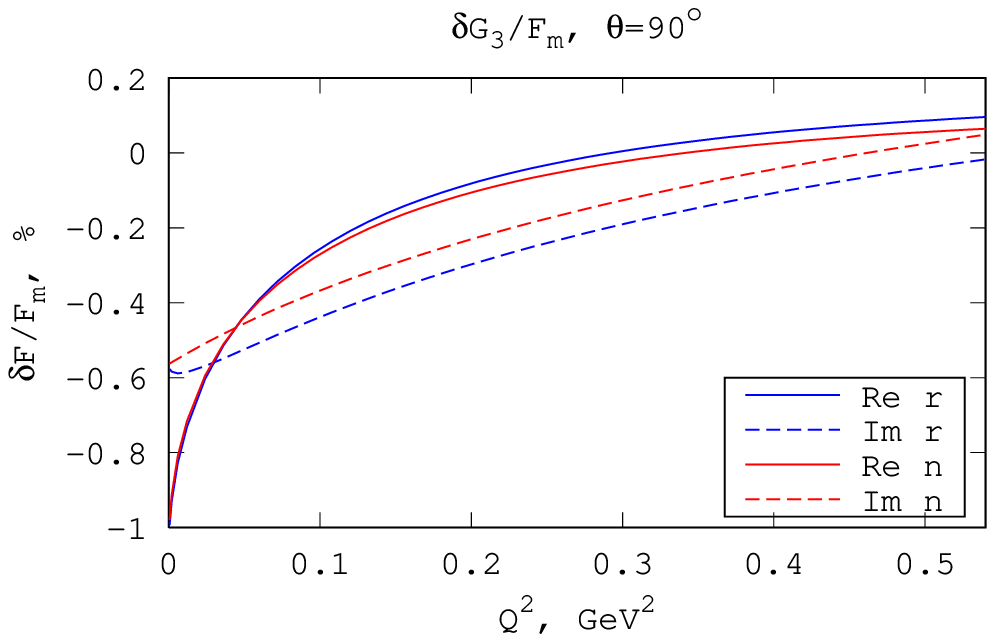}
\caption{The TPE amplitudes for proton target, $(\delta F_1+\delta\CG_3)/F_m$ (left), $\delta F_m/F_m$ (center) and $\delta\CG_3/F_m$ (right). Relativistic calculation (r), nonrelativistic (n).}\label{Fig:proton}
\end{figure}

Figures~\ref{Fig:proton} show the results of calculation of the TPE amplitudes for electron-proton scattering both using Eqs.(\ref{Int-k-begin}-\ref{Int-k-end}) (blue lines) and using relativistic calculation from Ref.~\cite{ourDisp} (red lines).
The scattering angle in the Breit system is fixed and equal $90^\circ$.

We see that, for the real part of the amplitude, NRA is rather good up to sufficiently large momentum transfers, $Q^2 \approx 0.5\GeV^2$, and the curves are almost indistinguishable at $Q^2 < 0.05\GeV^2$.
Surprisingly, the imaginary part of the amplitude start to disagree with NRA very early, at $Q^2 < 0.1\GeV^2$, though they coincide at $Q^2 \to 0$ as expected (at fixed scattering angle $Q^2 \to 0$ implies $E \to 0$).
This means that we should not rely on NRA when calculating observables depending on imaginary part, such as single-spin asymmetries, on nuclei.

The $^3$He nucleus has the same spin-parity $1/2^+$ as the proton, and the theory of TPE described here can be applied to it as well.
Of course, helium nucleus has %excited states and 
different internal structure, but if we (formally) consider elastic contribution only, the difference is just another values of mass and form factors. Thus we use $^3$He as another test for our formulae.

For the $^3$He nucleus, Eq.(\ref{Gamma}) should include a factor $Z=2$:
\be
  \Gamma_\mu(q) = Z \left\{ F_1(q^2) \gamma_\mu - \frac{1}{4M} [\gamma_\mu, \hat q] F_2(q^2) \right\}
\ee
and the normalization of form factors is then $F_1(0) = 1$, $F_2(0) = \mu M/(Z M_p) - 1 \approx -4.185$, where $\mu \approx -2.128$ is $^3$He nuclear magnetic moment, $M_p$ is proton mass and $M$ is nucleus mass. Accordingly, the expression (\ref{Gamma2}) for the TPE amplitude should be supplemented by a factor of $Z^2$. After this, all other relations, derived for the proton case, remain unchanged.

Usually, $^3$He electric and magnetic form factors $F_e$ and $F_m$ are normalized to unity, thus
\be
  F_1(q^2) = \frac{F_e(q^2) - \tfrac{q^2}{4M^2} \tfrac{\mu M}{Z M_p}  F_m(q^2)}{1 - \tfrac{q^2}{4M^2}}, \qquad
  F_2(q^2) = \frac{\frac{\mu M}{Z M_p} [F_m(q^2) - F_e(q^2)]}{1 - \tfrac{q^2}{4M^2}}
\ee

\begin{figure}[h]
\includegraphics[width=0.33\textwidth]{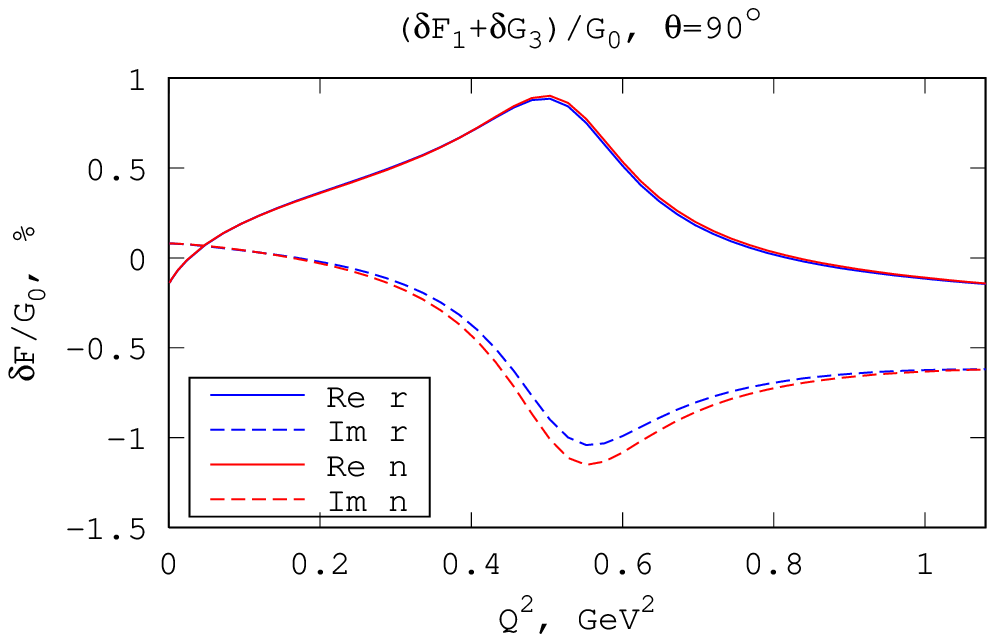}\hfill
\includegraphics[width=0.33\textwidth]{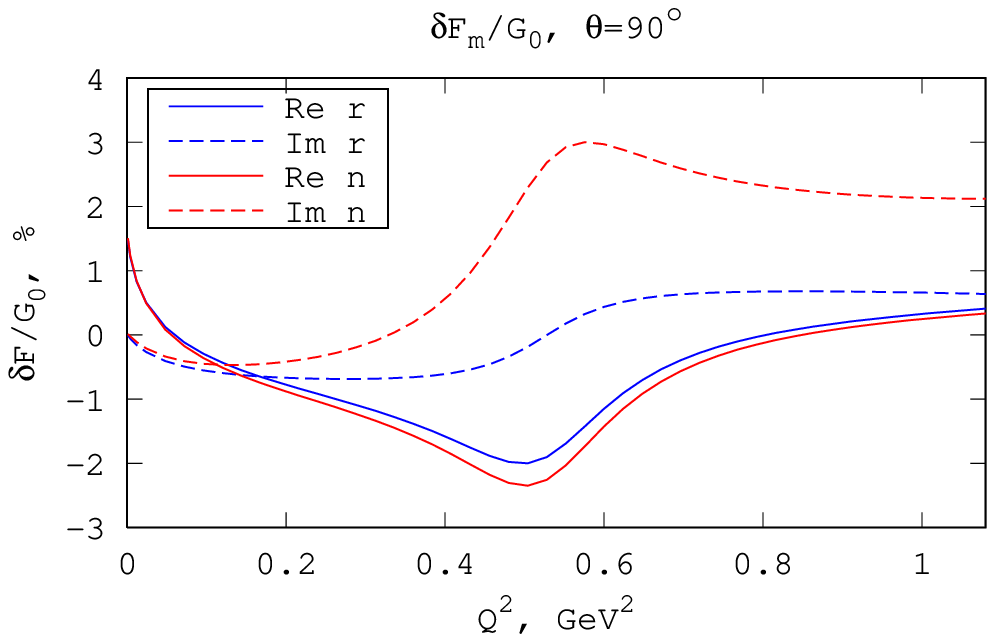}\hfill
\includegraphics[width=0.33\textwidth]{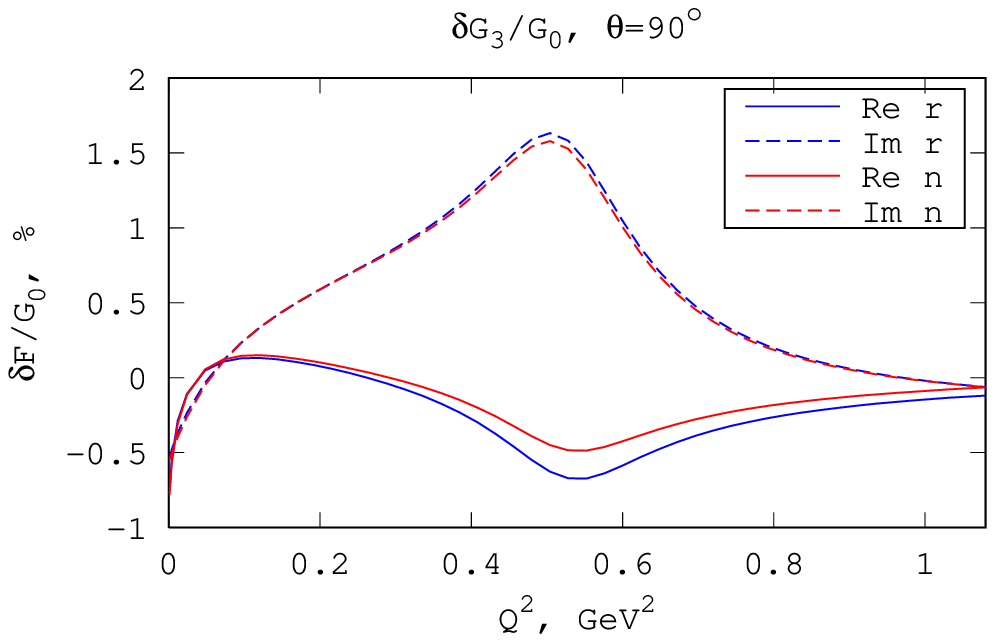}
\caption{The TPE amplitudes for $^3$He target, $(\delta F_1+\delta\CG_3)/G_0$ (left), $\delta F_m/G_0$ (center) and $\delta\CG_3/G_0$ (right). Relativistic calculation (r), nonrelativistic (n).}\label{Fig:He3}
\end{figure}

Figure~\ref{Fig:He3} is similar to Fig.~\ref{Fig:proton} but for the electron scattering off $^3$He, calculated with elastic form factors from Ref.~\cite{He3ff}.
Since both $^3$He for factors have zeros in the region of our interest, the TPE amplitudes are normalized not by $F_m$, but by always-positive quantity
\be
  G_0 = \sqrt{\sigma_0/\epsilon} = \left[F_e^2 - \frac{q^2 \mu^2}{4 \epsilon Z^2 M_p^2} F_m^2\right]^{1/2}
\ee
The figures for $^3$He show behaviour, similar to the proton case: real parts of the amplitude are well-described by NRA, but imaginary part (in particular, of the $F_m$ form factor) significantly differs from the relativistic result already at small $Q^2$, though goes to the correct limit at $Q^2\to 0$.
The bumps on the curves correspond to the zeros of the form factors.

\section{Conclusions}
In summary, we obtain formulae for the TPE amplitudes in the elastic electron scattering off the spin-1/2 hadron in the nonrelativistic approximation.
The amplitudes are expressed via three-fold integrals, which may be calculated analytically for the sum-of-poles form factor parameterization.

Numerical estimates for proton and $^3$He targets show that the real parts of the TPE amplitudes are well-described by the nonrelativistic approximation up to moderate $Q^2$ ($\sim 0.5\GeV^2$ for the proton), whereas the imaginary parts differ at much smaller $Q^2$, especially for the magnetic form factor.

This means that nonrelativistic approximation should be avoided in calculations of the observables which depend on imaginary part of the amplitude, such as single-spin asymmetries.

\appendix

\section{Deriving Eqs.(\ref{Int-k-begin}-\ref{Int-k-end})}\label{app:algebraic}

Let us transform the ''electron'' part of Eq.(\ref{Int-k-origin})
\bea
  \bar u' \gamma_0 (\hat k - \hat q_1) \gamma_0 u &=& 2k \bar u' \gamma_0 u - q_{1i} \bar u' \gamma_i u \\
  \ce_{ijk} q_{1k} \bar u' \gamma_0 (\hat k - \hat q_1) \gamma_j u &=& 2\cross{k}{q_1}_i \bar u' \gamma_0 u - i (q_{1i}q_{1j} - q_1^2\delta_{ij}) \bar u' \gamma_5 \gamma_j u \\
  \ce_{ijk} q_{1k} \bar u' \gamma_j (\hat k' + \hat q_1) \gamma_0 u &=& 2\cross{k'}{q_1}_i \bar u' \gamma_0 u - i (q_{1i}q_{1j} - q_1^2\delta_{ij}) \bar u' \gamma_5 \gamma_j u
\eea
which yields
\bea\label{result}
  \CM_2 = \frac{2\alpha^2}{\pi} \int d^3 q_1 
  \left\{ 2M F_1(q_1^2) F_1(q_2^2) S_0 \left( 2k \bar u' \gamma_0 u - q_{1i} \bar u' \gamma_i u \right) \frac{1}{(\vec k - \vec q_1)^2 - k^2 - i0} \right. \nonumber \\
          + 2i F_m(q_1^2) F_1(q_2^2) j_0 S_i \left[ \frac{\cross{k}{q_1}_i}{(\vec k - \vec q_1)^2 - k^2 - i0} + \frac{\cross{k'}{q_1}_i}{(\vec k' + \vec q_1)^2 - k^2 - i0} \right] \nonumber \\
          + \left. F_m(q_1^2) F_1(q_2^2) S_i (q_{1i} q_{1j} - q_1^2 \delta_{ij}) \bar u' \gamma_5 \gamma_i u \left[ \frac{1}{(\vec k - \vec q_1)^2 - k^2 - i0} + \frac{1}{(\vec k' + \vec q_1)^2 - k^2 - i0} \right]
  \right\}
\eea

Now we see that the answer can be expressed via the following integrals:
\bea
  I = \frac{1}{\pi^2} \int F_m(q_1^2) F_1(q_2^2) \frac{d^3 q_1}{(\vec k - \vec q_1)^2 - k^2 - i0} \label{Int-begin} \\
  I_i = \frac{1}{\pi^2} \int F_m(q_1^2) F_1(q_2^2) \frac{q_{1i} \, d^3 q_1}{(\vec k - \vec q_1)^2 - k^2 - i0} \\
  I_{ij} = \frac{1}{\pi^2} \int F_m(q_1^2) F_1(q_2^2) \frac{q_{1i} q_{1j} \, d^3 q_1}{(\vec k - \vec q_1)^2 - k^2 - i0} \label{Int-end}
\eea
and the same with the $F_m$ changed to $F_1$ (we denote the latter with the index $^{(1)}$).
These integrals depend on scalars $q^2$ and $k^2$, whereas $I_i$ and $I_{ij}$ also depend on vectors $\vec q$ and $\vec k_+ = 2\vec k - \vec q$ (which is more convenient than $\vec k$).

Analogous integrals with the denominators $(\vec k' + \vec q_1)^2 - k^2 - i0$ are obtained by changing $\vec k \to -\vec k'$. Under this operation scalars and the vector $\vec q$ remain unchanged, and the vector $\vec k_+$ changes its sign.

Let us write
\bea
  I_i &=& A k_{+i} + B q_{i} \\
  I_{ij} &=& C \delta_{ij} + D k_{+i} k_{+j} + E (k_{+i} q_j + q_i k_{+j}) + F q_i q_j
\eea
Substituting this into (\ref{result}), and taking into account that, as can be verified in a straightforward manner,
\be
 \bar u' \gamma_5 \vec \gamma u = \frac{i}{q^2} \left\{ \cross{k_+}{q} j_0 - 2k \cross{j}{q} \right\}
\ee
as well as
\bea
  \vec j \vec k_+ &=& 2k j_0 \\
  \cross{j}{q} &=& \frac{1}{k_+^2} \left\{ \vec k_+ (\vec k_+ \cross{j}{q}) + \vec k_+\vec j \cross{k_+}{q} \right\}
     = \frac{1}{k_+^2} \left\{ \vec k_+ (\vec k_+ \cross{j}{q} ) + 2k j_0 \cross{k_+}{q} \right\}
\eea
we obtain
\bea
  \CM_2 = 2\pi\alpha^2 \left\{ 4M k j_0 S_0 ( I^{(1)} - A^{(1)} ) + 2 i \cross{k_+}{q} \vec S j_0 \left(B - A + D - \frac{2C}{q^2} - F\right)
      + \frac{4ik}{q^2} \cross{j}{q} \vec S ( 2C + q^2 F ) \right\}
\eea
Comparing with (\ref{Mgeneric}), we see that
\bea
  \delta F_1 + \delta \CG_3 &=& \alpha k q^2 ( I^{(1)} - A^{(1)} ) \\
  \delta F_m              &=& 2\alpha k q^2 (2C/q^2 + F) \\
  \delta \CG_3            &=& - 2\alpha k q^2 ( 2C/q^2 + F + A - B - D)
\eea

Now, expressing the coefficients $A$-$D$ through the integrals (\ref{Int-begin})-(\ref{Int-end}),
and making %для удобства дальнейших вычислений
variable substitution $q_1 \to k'' = k-q_1$, we obtain equations (\ref{Int-k-begin})-(\ref{Int-k-end}).

\section{Calculation of the integrals}\label{app:integrals}

Expressing form factors as sums of simple poles
\be
  F_i(q^2) = \sum_n \frac{c_{in}}{q^2 + m_n^2}
\ee
we can reduce the integrals (\ref{Int-k-begin})-(\ref{Int-k-end}) to a linear combination of the integrals
\be
  J_{\la\mu}[X] = \frac{1}{\pi^2} \int \frac{X d^3 k''}{[q_1^2 + \la^2][q_2^2 + \mu^2][k''^2 - k^2 - i0]}
\ee
(where $\vec q_1 = \vec k-\vec k''$, $\vec q_2 = \vec k''-\vec k'$, and $X$ is some polynomial in components of the vector $\vec k''$),
and those, in turn, to the following 5 integrals:
\bea
&& J_{\la\mu}[1], \nonumber \\
&& J_{\la\mu}[k''^2 - k^2], \nonumber \\ 
%&& J_{\la\mu}[(k''^2 - k^2)(k''^2 - k^2 + \tfrac{\la^2+\mu^2+q^2}{2})], \nonumber \\
&& J_{\la}[1] \equiv J_{\la\mu}[q_2^2 + \mu^2], \nonumber \\
&& J_{\la}[\vec k''] \equiv J_{\la\mu}[\vec k''(q_2^2 + \mu^2)], \nonumber \\
&& J_{\la}[k''^2 - k^2] \equiv J_{\la\mu}[(k''^2 - k^2)(q_2^2 + \mu^2)]. \nonumber
\eea
They can be calculated analytically. To do this, one can, for example, integrate over angles, then expand the integration over $|k''|$ to the whole real axis and close the integration contour in the higher semi-plane.
The last integral contains an ultraviolet divergence,
which is eliminated by multiplying the integrand by $\frac{\La^2}{k''^2+\La^2}$, $\La \to \infty$.

Below we write down the formulae for these integrals which are needed in our calculations. The most complicated is the first one:
\be
  J_{\la\mu}[1] = \frac{i \pi^2}{\sqrt{R}} \ln \frac{k[(\mu+\la)^2+q^2]+i\la\mu(\mu+\la)+\sqrt{R}}{k[(\mu+\la)^2+q^2]+i\la\mu(\mu+\la)-\sqrt{R}}
\ee
where
\be
  R = k^2(\la^2+\mu^2+q^2)^2-(4k^2-q^2)\la^2\mu^2
\ee
Other integrals are simpler:
\be
  J_{\la\mu}[k''^2 - k^2] = \frac{i \pi^2}{q} \ln \frac{\la+\mu-iq}{\la+\mu+iq}
\ee
%\be
%  J_{\la\mu}[(k''^2 - k^2)(k''^2 - k^2 + \tfrac{\la^2+\mu^2+q^2}{2})] = \pi^2 (2\La-\mu-\la) 
%\ee
\be
  J_{\la}[1] = \frac{i \pi^2}{k} \ln \frac{\la-2ik}{\la}
\ee
\be
  J_{\la}[\vec k''] = \frac{i \pi^2 \vec k}{k^2} \left\{ i\la - k + \frac{2k^2+\la^2}{2k} \ln \frac{\la-2ik}{\la} \right\}
\ee
\be
  J_{\la}[k''^2 - k^2] = \frac{i \pi^2}{k} \La^2 \ln\frac{\La+\la-ik}{\La+\la+ik} \approx 2\pi^2(\La-\la)
\ee
Another useful forumae are
\be
  J_{\la\mu}[\vec k_+ \vec k''] = - \frac{1}{2} J_\la[1] - \frac{1}{2} J_\mu[1] + J_{\la\mu}[k''^2-k^2] + \frac{4k^2+\la^2+\mu^2}{2} J_{\la\mu}[1]
\ee
\be
  J_{\la\mu}[\vec q \vec k''] = \frac{1}{2} J_\la[1] - \frac{1}{2} J_\mu[1] + \frac{\la^2-\mu^2}{2} J_{\la\mu}[1]
\ee
\be
  J_{\la\mu}[(\vec k_+ \vec k'')^2] = -\frac{k_+^2}{4k^2} \left( J_\la[\vec k \vec k''] + J_\mu[\vec k' \vec k''] \right)
     + \frac{4k^2+\la^2+\mu^2}{2} J_{\la\mu}[\vec k_+ \vec k''] + \frac{\vec k_+^2}{2} J_{\la\mu}[k''^2-k^2]
\ee
\be
  J_{\la\mu}[(\vec q \vec k'')^2] = \frac{q^2}{4k^2} \left( J_\la[\vec k \vec k''] + J_\mu[\vec k' \vec k''] \right)
     + \frac{\la^2-\mu^2}{2} J_{\la\mu}[\vec q \vec k'']
\ee

\end{document}